\documentstyle{article}

\font\tenrm=cmr10
\font\tenit=cmti10
\font\elevenbf=cmbx10 scaled\magstep 1
\font\elevenrm=cmr10 scaled\magstep 1
\font\elevenit=cmti10 scaled\magstep 1

\newcommand{\be}{\begin{equation}}
\newcommand{\ee}{\end{equation}}
\newcommand{\e}{\epsilon}
\newcommand{\bt}{\beta}
\newcommand{\al}{\alpha}
\newcommand{\m}{\mu}
\newcommand{\n}{\nu}
\newcommand{\r}{\rho}
\newcommand{\s}{\sigma}
\newcommand{\lb}{\lambda}
\newcommand{\k}{\kappa}
\newcommand{\rz}{r_{\circ}}

\newcommand{\ti}{T_{\infty}}
\newcommand{\tl}{T_{loc}}
\newcommand{\th}{T_{H}}
\begin{document}
\begin{tabbing}
   \hspace{13.4cm} \=    \kill
  \>   \=  IFP-469 \+\+ \\
           July 1993
\end{tabbing}
\begin{center}
\vglue 0.6cm
{{\elevenbf ENERGY, TEMPERATURE, AND ENTROPY OF}\\
{\elevenbf BLACK HOLES DRESSED WITH QUANTUM FIELDS}\footnote{Based on a
talk given at the Fifth Canadian Conference on General Relativity and
Relativistic
Astrophysics, Waterloo, May, 1993.}}
\vglue 5pt
\vglue 1.0cm
{\tenrm James W. York, Jr. \\}
\baselineskip=13pt
{\tenit Institute of Field Physics and\\
Theoretical Astrophysics and Relativity Group\\
Department of Physics and Astronomy\\
The University of North Carolina\\}
\baselineskip=12pt
{\tenit Chapel Hill, North Carolina 27599-3255, USA\\}
\end{center}

\vglue 0.6cm
{\elevenbf\noindent 1. Introduction}
\vglue 0.4cm

\elevenrm
A deeper understanding of the thermal properties
of black holes than we presently have depends
to a large
degree on obtaining a firmer grasp of
the properties of the entropy. For such an understanding we must at least know
the basic relations among
entropy, energy, and temperature
of a black hole in thermal equilibrium with quantized matter fields. Limiting
attention
to spherical, uncharged (``Schwarzschild") holes,
we will find that the basic Bekenstein-Hawking relations have to be
generalized when the hole
is dressed by quantum fields. Though this fact
is not surprising, the corrections contain
surprises and are very instructive. My
purpose here is to discuss several aspects
of this problem and to display some concrete
results within the
framework of the semi-classical theory
of quantum fields in curved spacetime.$^{1}$ Two
key ideas that emerge are the following: (1)
The calculated thermodynamical entropy $\Delta S$
by which the quantum fields we consider augment
the usual Bekenstein-Hawking entropy$^{2\, , 3}$
is positive and monotonically increasing in
a suitable sense if and only if the back-reaction
of the quantum fields on
the spacetime geometry is taken into account.
(2) The total thermodynamical entropy within
a sphere of surface area $4 \pi r^{2}$, enclosing the black hole and fields,
can be
calculated
from the temperature
and the quasi-local energy determined at
radius $r$ without any reference to
asymptotic values of temperature or energy.

The quasi-local energy$^{4}$ also forms a
key part of a new fundamental approach to the statistical mechanics of
self-gravitating
systems. Such a statistical mechanics, that
when sufficiently developed should further illuminate black hole
thermodynamics, has
been introduced recently by means of a microcanonical functional
integral.$^{5}$
The later takes advantage of the property of
gravity by which the total energy of a system
may be known by
the behavior of the gravitational field
at its boundary. This behavior defines
the quasi-local energy. In this report,
however, the microcanonical functional
integral itself will not be used.

Progress in black hole thermodynamics
depends upon accounting for those features of gravity, such as its long-range
unscreened
nature, that partially contradict the usual thermostatistical assumptions. It
is
clear that the gravitational effects of the radiation that
would equilibrate a black hole
(or a star) prevent the existence of the
usual ``thermodynamical limit". Therefore we
should contemplate finite systems even
when back-reaction is being ignored,$^{6}$ in
order properly to set the stage for further
work. Thermodynamics will thus not look exactly
as it does in textbooks: it becomes more interesting. The generality and
simplifying
power that make thermodynamics a powerful
tool remain.

\vglue 0.6cm
{\elevenbf\noindent 2. Temperature and Energy}
\vglue 0.4cm

Consider a static spacetime ${\cal M}$.
(Stationary spacetimes are treated similarly
but involve further subtleties, not discussed
here, that have been treated in Refs. {\large $7$} and {\large $8$}.) We
learned from
Tolman$^{9}$
that quantities like temperature and
chemical potential, usually regarded
as intensive, in the presence of gravity are
no longer purely so. For example, consider
the temperature $T$. Then we have
as a consequence of the ``red shift" law that

\be
N\left(\vec{x}_{1}\right) T\left(\vec{x}_{1}\right)  =
N\left(\vec{x}_{2}\right) T\left(\vec{x}_{2}\right)
\label{redshift}
\ee
for any points $\vec{x}_{1}$ and $\vec{x}_{2}$
on a static time slice $t =$ constant.
Here $T\left(\vec{x}\right)$ denotes the
temperature measured locally by a static
observer at $\vec{x}$ and

\be
N\left(\vec{x}\right) = \left[ - g^{tt}\left( \vec{x}\right)\right]^{-1/2}
\label{lapse}
\ee
is the lapse function. If there is an ``asymptotically flat region at
spatial infinity" where $N \rightarrow 1$ then
there is a corresponding ``temperature at infinity" $\ti$. This is the
temperature usually meant,
for example, when one speaks of ``the Hawking temperature of a black hole".
However, as
we shall see, one can just as well, indeed
better, use a finite sphere rather than the asymptotic ``sphere at infinity" as
a reference locus for
the temperature of a star or a
black hole. (In the absence of spherical symmetry, other generally finite
reference surfaces
may prove simpler to
use. In principle, any two-surface enclosing
the system will suffice.)

Suppose we know the entropy $S$ as a
function of energy $E$, charge $Q$, and other conserved quantities. For
simplicity
let us
assume spherical symmetry
in what follows. Then, by definition,

\be
\bt (r) = \left( \frac{\partial S}{\partial E}\right)_{r, Q, \cdots}\, .
\label{inversetempA}
\ee
What energy is ``conjugate" to $\bt (r)$? If
$r = \infty$ is an asymptotically flat region,
then $E$ should be the ADM mass $M$ of the
system. More generally, however, $E$ is
the quasi-local energy obtained by a
geometrical analysis of the surface terms
belonging to the Hamiltonian
generating
the motion of the system bounded spatially
by the two-surface $B$, in this case $r =$ constant. As such, the
expression for $E$ is not
offered as a {\elevenit proposal} for
the quasi-local energy; it is dictated by the
action integral of the theory.$^{4}$ One
finds

\be
E(B) = \frac{1}{8\pi} \int_{B} \left( k - k_{0}\right)\sqrt{\s} d^{2}x\, ,
\label{qlenergyA}
\ee
where $\s_{ij}$ is the two-metric induced
on $B$ and $k = \s^{ij}k_{ij}$ is
the trace of the extrinsic curvature $k_{ij}$
of $B$ as embedded in the time slice $t =$
constant. (Units are chosen such that $G = c = k_{B} = 1$, but $\hbar \neq 1$.)
The term $k_{0}$
serves to define the zero of energy, which is
no more intrinsically fixed in general
relativity than it is in any other theory.
(The usual convention is that Minkowski
spacetime has zero energy.) A convenient
definition of $k_{0}$, when it can
be implemented,$^{4}$ is that it is the trace of
the second fundamental tensor of the
two-surface $B$ when the latter is
isometrically embedded in flat three-space.
This assures that $E$ becomes asymptotically the
ADM mass when the latter is defined.
In
any event, no physical result should depend
in an essential way on the zero of energy, and
that is true of the results reviewed in this
paper.

Consider a static spherically symmetric
spacetime in the usual coordinates,
such that $r$ is the ``areal" radius.
Temporarily ignore quantum fields and
back-reaction but let $r < \infty$. Then for
the Schwarzschild black hole,$^{6}$ using
Eq. (\ref{qlenergyA}), one finds

\be
E(r) = r - r\left(1 - \frac{2 M}{r}\right)^{1/2}\, ,
\label{schwarzenergy}
\ee
and, more generally, for the
Reissner-Nordstr\"{o}m black hole$^{10}$

\be
E(r) = r - r\left( 1 - \frac{2 M}{r} + \frac{Q^{2}}{r^{2}}\right)^{1/2}\, .
\label{reissnordenergyA}
\ee
Note that Eq. (\ref{reissnordenergyA})
implies

\be
M = E - \frac{E^{2}}{2 r} + \frac{Q^{2}}{2 r}\, .  \label{reissnordmass}
\ee
The second and third terms on the right
have the expected signs for gravitational and electrostatic
binding energies associated
with constructing
a shell of radius $r$, energy $E$, and
charge $Q$. But Eq. (\ref{reissnordmass}) just helps us build
a ``word picture" to make us feel
more comfortable with the energy $E$.
Ref. {\large $4$} contains further heuristic discussions along
these lines. However,
the important expression is Eq. (\ref{reissnordenergyA}), which
can be expressed without using the ADM mass
$M$. In fact, $M$ has no essential role as mass
in this problem. It is rather just a
parameter
perhaps better expressed in terms of the gravitational radius of the
hole,

\be
r_{+} = M + M \left( 1 - \frac{Q^{2}}{M^{2}}\right)^{1/2}\, . \label{rplus}
\ee
In place of Eq. (\ref{reissnordenergyA}) we find

\be
E(r) = r - r \left[\left(1 - \frac{r_{+}}{r}\right)\left(
1 - \frac{Q^{2}}{r r_{+}}\right)\right]^{1/2}\, . \label{reissnordenergyB}
\ee
Given that the Bekenstein-Hawking entropy
is $S_{BH} = \pi r_{+}^{2} \hbar^{-1}$, we
find

\be
\left(\frac{\partial S_{BH}}{\partial E}\right)_{r, Q} = \bt (r) = \left[
\frac{4 \pi r_{+}}{\hbar} \left( 1 -
\frac{Q^{2}}{r_{+}^{2}}\right)^{-1}\right]\left[\left(
1 - \frac{r_{+}}{r}\right)\left(1 - \frac{Q^{2}}{r r_{+}}\right)\right]^{1/2}\,
{}.
\label{inversetempB}
\ee
This is the inverse Hawking temperature, blue-shifted from
infinity to $r$.
Note that the result is independent of
the choice of the zero-point for $E$
(represented by the first term on the
right in Eq. (\ref{reissnordenergyB})). The inverse of the first
factor in the rectangular
brackets on the right of  Eq. (\ref{inversetempB}) is just $\k \hbar / 2 \pi$,
where $\k$ is the surface gravity of the event horizon.
The second factor in the rectangular
brackets on the right of Eq.
(\ref{inversetempB}) is the lapse
function $N(r)$. The result for $\bt (r)$
can be expressed in terms of $r$, $E$,
and $Q$ by
using Eq. (\ref{reissnordmass}) and Eq. (\ref{rplus}) in
Eq. (\ref{inversetempB}).
The conclusion is that the energy $E$ as defined by the
surface term Eq. (\ref{qlenergyA})
from the behavior of the gravitational
field on the two-boundary $B$ is both
the energy defined by the total
Hamiltonian of
the system$^{4}$ {\elevenit and}
the total internal energy of the system
in the sense of thermodynamics. Identification
of these quantities is a
significant step in the unification
of gravitation and thermodynamics that was anticipated when the
Hawking effect
was discovered.

\vglue 0.6cm
{\elevenbf\noindent 3. Quantum Stress-energy Tensors and Back-reaction}
\vglue 0.4cm

A black hole can exist in thermodynamical equilibrium provided that it is
surrounded by radiation with a suitable
distibution of stress-energy. In
the semi-classical approach, such
radiation is characterized by the
expectation value of a stress-energy tensor
obtained by renormalization of a quantum field
on the classical spacetime geometry
of a black hole. One can use such a
stress-energy tensor as a source in the semi-classical Einstein equation,

\be
G^{\m}_{\n} = 8 \pi \langle T^{\m}_{\n}\rangle_{renormalized}\, ,
\label{einsteineq}
\ee
to calculate the change effected
by the stress-energy tensor in the black hole's spacetime metric. This is
the ``back-reaction" problem associated with the
spacetime geometry of a black hole in equilibrium.

We shall see, from the properties of the renormalized stress-energy
tensors
we employ and of the semi-classical Einstein equation, that we can obtain
accurate fractional corrections to the metric only in $O(\e )$, where $\e =
\hbar M^{-2}$, $M_{Pl} =
\hbar^{1/2}$ is the Planck
mass and $M$ is the mass
of the black hole. Because the
usual black hole entropy $S_{BH} = ( 4 \pi M^{2} ) \hbar^{-1} = O ( \e^{-1} )$,
corrections
to $S_{BH}$ can be obtained in $O (\e^{0} ) = O(1)$ from the fractional
corrections
of $O(\e)$ in the metric. It turns out that these corrections are of the same
order
as the naive
flat space radiation entropy $(4/3)a \th^{3} V$, where $a = (\pi^{2}/15
\hbar^{3} )$,
$\th = \hbar(8 \pi M)^{-1}$ is the uncorrected Hawking
temperature of a Schwarzschild black hole, and $V$ is the flat space volume.
{}From this
fact alone it follows that the back-reaction
cannot be ignored.

Stress-energy tensors renormalized on a Schwarzschild background have been
obtained in exact form for conformal scalar fields and for $U(1)$ gauge fields,
respectively, by Howard$^{11}$ and by Jensen and
Ottewill.$^{12}$ Both results can be
written in the form

\be
\langle T^{\m}_{\n} \rangle_{renormalized} = \langle T^{\m}_{\n}
\rangle_{analytic} +
\left( \frac{\hbar}{\pi^{2} (4M)^{4}}\right) \Delta^{\m}_{\n}\, ,
\label{stress-energy}
\ee
where the analytic piece, in the
case of the conformal scalar field, was given by Page.$^{13}$
The term $\Delta^{\m}_{\n}$ is
obtained from a numerical evaluation of a mode sum. The numerical piece is
small compared to
the analytic piece, and we do not include it in the calculations in this paper.
This does
not change any of the results qualitatively because
both pieces separately obey the required regularity and consistency conditions.
The
analytic piece has the exact
trace anomaly in both cases. The
stress-energy tensors have vanishing covariant divergence on the Schwarzschild
background. They represent the stress-energy distribution required to
equilibrate the
black hole with its own Hawking radiation.
Each satisfies $\langle T^{t}_{t}\rangle = \langle T^{r}_{r}\rangle$ at the
horizon $r = 2M$, which is required for regularity of the spacetime
geometry.$^{13}$ Each has the
asymptotic form of a flat spacetime radiation stress-energy tensor at the
uncorrected Hawking temperature at infinity of
an ordinary Schwarzschild black
hole, denoted by $\th = \hbar (8 \pi M)^{-1}$.
The cases worked out so far$^{14}$ include
the conformal scalar field,
the massless spin $1/2$ field,$^{15}$ and the $U(1)$ vector field. Only the
latter will
be displayed here.

Dropping the angular brackets and
displaying the analytic piece, one has for
the $U(1)$ vector field, with $w = 2M/r$,

\be
T^{t}_{t} = - \frac{1}{3} a \th^{4} \left( 3 + 6w + 9w^{2} + 12w^{3} - 315w^{4}
+
78w^{5} - 249w^{6}\right)\, ,
\ee
\be
T^{r}_{r} = \frac{1}{3} a \th^{4} \left( 1 + 2w + 3w^{2} - 76w^{3} + 295w^{4} -
54w^{5} +
285w^{6}\right)\, ,
\ee

\be
T^{\theta}_{\theta} = T^{\phi}_{\phi} = \frac{1}{3} a \th^{4} \left( 1 + 2w +
3w^{2} + 44w^{3} -
305w^{4} + 66w^{5} - 579w^{6}\right)\, .
\ee
Note that $T^{r}_{r} > 0$ and
that the energy density $- T^{t}_{t}$ is negative in the vicinity of the
event horizon, thus
violating the weak energy
condition. The energy density is
negative from $r =2M$ to $r \approx 5.14M$. The dominant energy condition is
also
violated in a region surrounding
and bordering the horizon. It
is convenient in what follows to write

\be
\frac{1}{3} a \th^{4} = \frac{\e}{48 \pi K M^{2}}\, ,
\ee
where $K = 3840\pi$.

We obtain fractional corrections $h^{\al}_{\n}$ to the metric by setting

\be
g_{\m\n} = \hat{g}_{\m\al}\left(\delta^{\al}_{\n} + \e h^{\al}_{\n}\right)
\ee
in the semi-classical Einstein
equation Eq. (\ref{einsteineq}), where $\hat{g}_{\mu\nu}$ is the uncorrected
Schwarzschild metric. We work in
linear order in $\e$ as required by $\hat{\nabla}_{\m} T^{\m}_{\n} = 0$ and
$\hat{\nabla}_{\m} (\delta G^{\m}_{\n}) = 0$, where $\delta G^{\m}_{\n}$ is the
Einstein operator linearized on a
background satisfying $\hat{G}^{\m}_{\n} = 0$. The corrected geometry will be
taken
to be static and spherically
symmetric. Working out the
equations as in Refs. {\large $16$}
and {\large $17$}, we find the corrected
metric can be written as

\be
ds^{2} = -\left( 1 - \frac{2m(r)}{r}\right)\left(1 + 2\e\bar{\r}(r)\right)
dt^{2} + \left(1 - \frac{2 m(r)}{r}\right)^{-1} dr^{2} + r^{2} d\omega^{2}\, ,
\label{correctedmetric}
\ee
where $d\omega^{2}$ is the standard metric of a normal round unit sphere.
To obtain $m(r)$ and $\bar{\r}(r)$ requires only simple radial integrals
involving $T^{t}_{t}$
and $T^{r}_{r}$. The angular components enter linearized Einstein equations
that hold automatically by virtue of $\hat{\nabla}_{\m} T^{\m}_{\n} = 0$ in a
static spherical geometry.

The mass function $m(r)$ has the form

\be
m(r) = M\left( 1 + \e \m(r) + \e\, C K^{-1}\right)\, , \label{massfuncA}
\ee
with

\be
\m(r) = \frac{1}{\e M} \int^{r}_{2M} \left( - T^{t}_{t}\right) 4 \pi
\tilde{r}^{2}\,
d\tilde{r}\, ,   \label{muintegral}
\ee
so $\m(r)$ vanishes at the horizon.
In Eq. (\ref{massfuncA}), $C$ is an undetermined integration constant that
inspection
of Eq. (\ref{correctedmetric})
shows is to be absorbed into $M$ to obtain a renormalized mass for the black
hole.
Thus, setting $g^{rr} = 0$ shows that $r = 2m = 2M(1 + \e\, C K^{-1}) =
2M_{renormalized}$
locates the event horizon. Note that, to the order we are working, we can write
$m(r) =
M(1+ \e\, C K^{-1})(1 + \e \m(r)) \equiv M_{ren}(1 + \e \m(r))$. The
renormalized mass will not be
distinguished notationally
from the original Schwarzschild mass $M$
in what follows, as the
bare Schwarzschild mass has no physical meaning in the back-reaction problem.
Therefore, we
write

\be
m(r) = M \left( 1 + \e \m(r) \right) \equiv M + M_{rad}(r)  \label{massfuncB}
\ee
where, using Eq. (\ref{muintegral}),
we see that $M_{rad} = \e\, M \m$ is the usual expression for the effective
mass of a
spherical source. One finds

\be
K \m = \frac{2}{3}w^{-3} + 2w^{-2} + 6w^{-1} - 8ln(w) + 210w - 26w^{2} +
\frac{166}{3}w^{3} - 248\, . \label{Kmurelation}
\ee
In Eq. (\ref{Kmurelation}), we
note that the first term on the right, multiplied by $\e M K^{-1}$, gives
the naive flat-space value $a \th^{4} V$ for
radiation energy.

The metric is completed by a determination of $\bar{\r}$ which, like $\m$,
can be found from an elementary integration. Defining

\be
K \bar{\r} \equiv K \r + k\, ,
\ee
where $k$ is constant of
integration (not the same as the $k$ in Eq. (\ref{qlenergyA})), we have

\be
\r = \frac{1}{\e} \int^{r}_{2 M} \left( T^{r}_{r} - T^{t}_{t}\right)\left(
\tilde{r} -
2 M\right)^{-1} 4 \pi \tilde{r}^{2} d \tilde{r}\, . \label{rhointegral}
\ee
We find

\be
K \r = \frac{2}{3}w^{-2} + 4w^{-1} - 8 ln (w) + \frac{40}{3}w + 10w^{2} +
4w^{3} - 32\, ,
\ee
with $\r (1) = 0$ at $w = 1$.

Because the radiation stress-energy
tensors are asymptotically constant, it
is clear that the system composed of black hole plus equilibrium radiation
must be put in a finite ``box". Otherwise,
the fractional corrections $\e\, h^{\al}_{\n}$ to the metric would not remain
small for
sufficiently
large radius. Physically this means that the radiation in a box that is too
large would collapse onto
the black hole, producing a
larger one. Hence, we must chose the radius $\rz$ of the box such that it is
less than the second
positive root $r_{\ast}$ for $r$ in $g^{rr} = 0$ (the first zero corresponds to
the horizon $r =2 M$).
We shall also assume that the box radius $\rz$ is sufficiently large that the
stress-energy tensors we employ, which were constructed
for infinite asymptotically flat spacetime, are a good approximation. Clearly,
a
finite radius would cut out
some of the radial modes that were used in these calculations. However, if
$\rz$
is somewhat greater then the longest wavelength characteristic of Hawking
radiation,
which in turn is
associated with the least-damped quasi-normal mode of lowest angular momentum
for the
field in question, then this effect should be negligible. This wavelength
$\lb_{\ast}$ is about $42 M$
for the conformal scalar field and is
smaller for the higher-spin massless fields. Also, if $\rz > \lb_{\ast}$, then
the
explicit nature of the walls of the box ({\elevenit e.g.}, adiabatic {\elevenit
versus} diathermic)
should not be important. For these reasons we shall assume throughout the
remainder
of this work that $\lb_{\ast} < \rz < r_{\ast}$. (Of
course, one
must also assume that $M \stackrel{>}{\sim} M_{Pl}$, in any treatment based on
Eq. (\ref{einsteineq}).)

One convenient way to fix the
constant $k$ is to impose a microcanonical boundary condition.$^{16}$ We fix
$\rz$ and
imagine placing there an ideal massless perfectly reflecting wall. Outside
$\rz$, we then have an
ordinary Schwarzschild spacetime

\be
ds^{2} = - \left( 1 - \frac{2 m(\rz)}{r}\right) dt^{2} + \left( 1 -
\frac{2 m(\rz)}{r}\right)^{-1} dr^{2} + r^{2} d\omega^{2}\, ,
\label{schwarzmetric}
\ee
for $r \geq \rz$. Continuity of the three-metric induced by the metrics Eq.
(\ref{correctedmetric})
and Eq. (\ref{schwarzmetric}) on
the world tube $r = \rz$ fixes the constant $k$ in $\bar{\r}$ by the relation

\be
k = - K \r\left( \rz\right)\, .  \label{krelation}
\ee
There are finite discontinuities
in the extrinsic curvature of the world
tube $r = \rz$,$^{16}$ but these, and other properties of the box wall, are of
no interest in the present analysis, as we
argued above. The space time geometry, including back-reaction, is now
completely determined by Eq.
(\ref{schwarzmetric}) for $r \geq \rz$, and for $r \leq \rz$ by Eq.
(\ref{correctedmetric}) and
Eq. (\ref{krelation}).

\vglue 0.6cm
{\elevenbf\noindent 4. Temperature and Entropy}
\vglue 0.4cm

If we release a small packet of energy from a closed box containing a black
hole through a
long thin radial tube, it will undergo a
red-shift and
approach the asymptotic temperature

\be
\ti = \frac{\k_{H} \hbar}{2 \pi}\, , \label{infinitytempA}
\ee
where $\k_{H}$ is the surface
gravity of the event horizon. For an ordinary Schwarzschild black hole
(ignoring radiation), one
finds $\k_{H} = (4 M)^{-1}$ and $\ti = \th = \hbar(8\pi M)^{-1}$ . However, the
stress-energy of the
radiation changes the surface gravity of the horizon to

\be
\k_{H} = \left. \frac{1}{4M} \left[ 1 + \e \left(\bar{\r} - \m\right) +
8 \pi r^{2} T^{t}_{t}\right] \right|_{r = 2M}\, ,   \label{surfacegravity}
\ee
as a straightforward calulation shows.$^{16}$ With the microcanonical boundary
conditions, we can use Eq. (\ref{krelation}) to obtain from Eq.
(\ref{infinitytempA})
and Eq. (\ref{surfacegravity})

\be
\ti = \frac{\hbar}{8 \pi M} \left[ 1 - \e \r\left( \rz\right) +
\e\, n K^{-1}\right]\, ,  \label{infinitytempB}
\ee
where $n$ takes the value $304$ for the vector field. (It has other values
for other fields.) The local temperature at the boundary of the box is
obtained by blue-shifting Eq. (\ref{infinitytempB}) from infinity back to
$\rz$.
We find from

\be
\tl = \ti \left[ - g_{tt} \left( \rz\right)\right]^{-1/2}\, ,
\label{localtempA}
\ee
that

\be
\tl\left( \rz\right) = \frac{\hbar}{8 \pi M} \left[ 1 - \e  \r \left( \rz
\right) + \e\, n K^{-1} \right] \left[ 1 -
\frac{2 m\left(\rz\right)}{\rz}\right]^{- 1/2}\, .   \label{localtempB}
\ee
The temperature $\tl$, unlike $\ti$, is actually {\elevenit independent} of the
boundary condition that
determines the constant $k$, as
explained in detail in Ref. {\large $16$}. Indeed, it can be readily verified
that $k$ cancels
out in $O(\e )$ in the expression Eq. (\ref{localtempA}) for $\tl$. {\elevenit
Either}
measure of temperature, $\ti$ or $\tl$, can be used to calculate the same
entropy in
conjunction with an appropriate measure of energy. This is quite important: it
means
that the specific boundary condition chosen does {\elevenit not} affect the
calculated entropy, as we shall
see below.

One way to calculate the entropy is as follows. Fix the radius $\rz$ of a
closed box. The measure of energy
in the box conjugate to the asymptotic
inverse temperature $\bt_{\infty} \equiv \ti^{-1}$ is then the ADM mass
$m(\rz)$ determined
at spatial infinity. The first law of thermodynamics for slightly differing
equilibrium configurations
with the same areal radius tells us
that

\be
dS = \bt_{\infty} dm \hspace{1cm} \left(d\rz = 0\right)\, ,
\ee
where $S(\rz)$ is the total entropy in the box. By this method we seem to
obtain only
the total entropy $S(\rz)$ rather than the distribution of entropy in the given
box, $S(r)$,
for $r \leq \rz$, where $S(r)$ denotes the total
entropy inside the radius $r$. However, the
latter can be obtained by using the quasi-local energy $E$,$^{4}$ which for
static
spherical metrics like those treated here is given by

\be
E(r) = r - r\left[ g^{rr}(r)\right]^{1/2}\, , \label{qlenergyB}
\ee
with $g^{rr}(r)$ determined by the metric for $r \leq \rz$. This energy, unlike
$m$,
does not depend on asymptotic flatness in its definition, nor even on
the existence of an
asymptotically flat region.$^{4}$ Furthermore, even the ``normalization" of the
zero
of energy$^{4}$ that is incorporated in $E$ as given in Eq. (\ref{qlenergyB})
does not affect
the calculated entropy, as it certainly
should not. (Recall that this ``normalization" is intended to make $E$ approach
the ADM mass
in an asymptotically flat region, if such a region exists.) Similarly, the
inverse local temperature $\bt (r) \equiv \tl^{-1} (r)$, $r \leq \rz$, is
independent of the boundary condition as metioned
above. Hence, {\elevenit the value of the entropy depends neither on the zero
of energy nor the existence of an
asymptotic region.}

Therefore, to obtain $S(r)$, in place of Eq. (\ref{localtempA}) we can write

\be
dS = \bt dE\,\,\,\, \left( dr = 0,\,\, r \leq \rz\right)\, .  \label{firstlawA}
\ee
Choosing $M$ and $r$ as independent variables, and fixing $r$, we can readily
integrate Eq. (\ref{firstlawA}) to obtain $S$ up to a function of $r$ and a
constant. From
Eq. (\ref{localtempB}) we have

\be
\bt (r) = \frac{8 \pi M}{\hbar} \left[ 1 + \e\r (r) - \e\, n
K^{-1}\right]\left[ 1 -
\frac{2 m (r)}{r}\right]^{1/2}\, , \label{inversetempC}
\ee
and from Eq. (\ref{massfuncB}) and Eq. (\ref{qlenergyB}), holding $r$
fixed,

\be
dE = \left[ 1 -\e\m + \e M\frac{\partial \m}{\partial M}\right]\left[1 -
\frac{2m(r)}{r}\right]^{-1/2} dM\, .  \label{dE}
\ee
One can see directly for any $r \leq \rz$ that $\bt_{\infty} dm = \bt dE$
where, of course, one replaces $\rz$ by $r$ in the
formulas for $\bt_{\infty}$ and $m$ to establish this result. This equality
means that we can calculate $S(r)$ for any $r \leq \rz$ . The key point of
this discussion is that Eq. (\ref{firstlawA}) is independent of the boundary
conditions (does not
depend
on $k$). This means that the $\Delta S(r)$ calculated does not depend
on whether there is empty
space just outside $r$ or more radiation. The result will hold for any $r$ such
that $2 M \leq r < \rz$.

Observe that from the fractional changes of $O(\e )$ in the metric, which
affect the surface
gravity and temperature in this order, we
are able to calculate from Eq. (\ref{firstlawA}) departures of $O(\e^{0}) =
O(1)$ from the usual
black hole entropy $S_{BH} = (4 \pi M^{2})\hbar^{-1} = 4 \pi \e^{-1}$. But in
fact all
of the corrections to the entropy are of the {\elevenit same} order as the
naive flat-space entropy itself:

\be
\frac{4}{3}a\th^{3}V =
\frac{4}{3}\left(\frac{\pi^{2}}{15\hbar^{3}}\right)\left(\frac{\hbar}{8 \pi
M}\right)^{3}\left(\frac{4}{3}\pi
r^{3}\right) = \frac{8\pi}{K}\left(\frac{8}{9} w^{-3}\right) = O(1) \times
w^{-3}\, . \label{flatentropy}
\ee
The $\hbar$'s in Eq. (\ref{flatentropy})
cancel out, leaving only a function
of $w = 2 M r^{-1}$. Combining Eq. (\ref{inversetempC}) and Eq. (\ref{dE})
yields

\be
dS = \frac{8 \pi M}{\hbar} dM + 8 \pi \left[w^{-1}(\r - \m) + \frac{\partial
\m}{\partial w} -
n K^{-1} w^{-1}\right] dw\, , \label{firstlawB}
\ee
with $dr =0$. Integration of Eq. (\ref{firstlawB}) gives an expression of the
form

\be
S = \frac{4 \pi M^{2}}{\hbar} + \Delta S(w) +
f\left(\frac{r}{\hbar^{1/2}}\right)\, ,\,\,\,\,\left(1 \leq w \leq w_{\circ} =
2 M / \rz\right)
\label{correctedentropy}
\ee
where the first term is the usual Bekenstein-Hawking expression $S_{BH}$ for
the black hole
entropy, the second term is a function of $w$
determined up to an additive integration constant by the second term on the
right of Eq.
(\ref{firstlawB}), and $f$ is a dimensionless function of $r$ that does not
depend on $M$.
The appearance of the function $f$ in Eq. (\ref{correctedentropy}) can be
understood as follows. Since our problem involves the three mass
or length scales $M_{Planck} = \hbar^{1/2}$, the mass of the black hole, M,
and the radius $r \leq \rz$, there are, for a given $r$, three relevant
dimensionless
parameters one can define, namely, $\e = \hbar M^{-2}$, $w = 2 M/r$ and
$r/\hbar^{1/2}$. However, the first
two terms on the right of Eq. (\ref{correctedentropy}) depend only on $\e$ and
$w$, respectively. Thus, if the
entropy $S$ depends on $r/\hbar^{1/2}$, it can only do so through a separate
function of this parameter.

Let us first dispose of the dimensionless function $f$, which clearly can
depend only
on $(r/\hbar^{1/2})$, where $\hbar^{1/2}$ is the Planck length in our units. It
seems that such a
term could only arise in a theory taking quantum gravity into explicit account
because
the semi-classical theory has incorporated the dimensionless terms involving
$\hbar/M^{2}$ and $2 M/r$.
(Of course, quantum gravity
could modify terms of these latter two types quantitatively.) On dimensional
grounds,
therefore, we take $f = 0$ in the semi-classical theory. The possibility of an
additive
constant will be discussed when
we treat $\Delta S$ below.

In considering $\Delta S$, which will be
given explicity below, we first note the
significant property that

\be
\frac{\partial (\Delta S)}{\partial w} = 8 \pi \left[w^{-1}(\r - \m) +
\frac{\partial \m}{\partial w} - n K^{-1} w^{-1}\right] \label{entropyderivA}
\ee
{\elevenit vanishes at the horizon} $w = 1$. Therefore, for a fixed black hole
mass $M$, the derivative with respect to $r$ of $\Delta S$ vanishes at the
horizon.
Thus $\Delta S$ has a local extremum with respect to $r$ at the horizon. This
result
follows from several general features that will be
enjoyed by {\elevenit all} regular renormalized stress-energy tensors on the
Schwarzschild background and the back-reactions they
induce, not just the presently
known cases. First, $\m$ vanishes at the horizon by virtue of the black hole's
mass having been
suitably renormalized. Second, $\r$ vanishes at the horizon, as follows from
Eq. (\ref{rhointegral}) and the regularity condition $T^{t}_{t} = T^{r}_{r}$ at
the horizon.
More precisely, we have that

\be
\lim_{w \rightarrow 1^{+}} \left(\frac{T^{t}_{t} - T^{r}_{r}}{1 - w}\right)
\ee
exists. Third, the last two terms on the
right side of Eq. (\ref{firstlawB}) add to zero at the horizon because there
the Hamiltonian
constaint $(G^{t}_{t} - 8 \pi T^{t}_{t} = 0)$ holds. Furthermore, note that if
the fractional
effects of $O(\e )$ in the temperature induced by the back reaction were
neglected, the
derivative of Eq. (\ref{entropyderivA}) would not vanish at the horizon.

Is the local extremum of $\Delta S$ at the
horizon a local minimum? To answer this we
calculate

\be
\frac{\partial^{2}(\Delta S)}{\partial w^{2}} = 8 \pi \left[ - w^{-2}(\r - \m)
+
w^{-1}\left(\frac{\partial \r}{\partial w} - \frac{\partial \m}{\partial
w}\right) +
\frac{\partial^{2} \m}{\partial w^{2}} + n K^{-1} w^{-2}\right]\, ,
\ee
which becomes, at the horizon $w = 1$,

\be
\left. \frac{\partial^{2}(\Delta S)}{\partial w^{2}}\right|_{w = 1} = 8 \pi
\left.\left(\frac{\partial \r}{\partial w} +
\frac{\partial^{2} \m}{\partial w^{2}}\right)\right|_{w = 1}
\label{entropy2derivA}
\ee
or, equivalently, with $M$ fixed,

\be
\left. \frac{\partial^{2}(\Delta S)}{\partial r^{2}}\right|_{r = 2M} =
\left.\frac{32 \pi^{2} M^{2}}{\hbar}\left[4M\frac{\partial \left( -
T^{t}_{t}\right)}{\partial r} -
8 T^{r}_{r} - \left(\frac{T^{r}_{r} - T^{t}_{t}}{1 -
2M/r}\right)\right]\right|_{r = 2M}\, .
 \label{entropy2derivB}
\ee
Hence we need only examine the stress-tensors.
In all the cases treated
so far$^{14}$ (conformal scalar, vector, massless fermion),  Eq.
(\ref{entropy2derivA})
and Eq. (\ref{entropy2derivB}) are positive so that $\Delta S$ takes a local
mimimum
with respect to the radius at the horizon.
This suggests, but does not prove, that $\Delta S$ is non-negative.

The local minimum of $\Delta S$ at the
horizon and the fact the $S_{BH}$ in the expression Eq.
(\ref{correctedentropy})
for the total entropy $S$ contains the {\elevenit renormalized}
mass $M$ of the hole motivate the choice of the
remaining additive constant in $\Delta S$, which can only be a pure number,
to be such that $\Delta S = 0$ at $w = 1$. For $w = 1$, with no ``room" for
the fields to contribute anything further, one then obtains only
the Bekenstein-Hawking entropy $(1/4)A_{H}\hbar^{-1}$, as would be expected.
With the choice $\Delta S(w = 1) = 0$, we obtain for
the vector field

\be
\Delta S = \frac{8 \pi}{K} \left(\frac{8}{9}w^{-3} + \frac{8}{3}w^{-2} +
8w^{-1} - 96ln(w) +
\frac{40}{3}w - 8w^{2} + \frac{344}{9}w^{3} - \frac{496}{9}\right)\, .
\ee
(Positive entropy $\Delta S$ for the conformal scalar field was first obtained
in Refs. {\large $18$} and {\large $19$}.) In this expression, the naive
flat-space
radiation entropy term Eq. (\ref{flatentropy}) appears as the
first term on the right. $\Delta S$ is
positive for $1 \geq w \geq w_{\circ} \geq w_{\ast} = 2Mr_{\ast}^{-1}$ and
vanishes at $w = 1$. Hence, in that it is positive, it is amenable to arguments
relating thermodynamical and statistical entropy. It is not heretofore been
evident
that this desirable feature would be present in the semi-classical theory. The
reader can
verify, by omitting the back-reaction terms in the inverse
temperature Eq. (\ref{inversetempC}),
that not only is the vanishing slope of $\Delta S$ at $w = 1$ lost, but
also that the value of the resulting ``$\Delta S$", normalized as above, is no
longer positive for the range $1 \geq w \geq w_{\circ}$ .In this fundamental
sense, we
conclude that the back reaction, however small quantitatively in
its effects on the metric near a black hole,
can never be regarded as negligible.
\vglue 0.5cm
{\elevenbf\noindent 5. Acknowledgements \hfil}
\vglue 0.4cm
I thank G. L. Comer, D. Hochberg, and T. W.
Kephart for collaboration in part of this work and for helpful discussions.
This research
was supported
by National Science Foundation grants PHY-8407492 and PHY-8908741.
\vglue 0.5cm
{\elevenbf\noindent 6. References}
\vglue 0.4cm
\elevenrm
\begin{list}{\arabic{enumi}.}{\usecounter{enumi}\setlength{\itemsep}{3pt}
\settowidth{\labelwidth}{2em}\sloppy}

\item N. D. Birrell and P. C. W. Davies, {\elevenit Quantum Fields in Curved
Space} (Cambridge
University Press, Cambridge, 1982).

\item J. D. Bekenstein, {\elevenit Phys. Rev.} {\elevenbf D7} (1973) 2333.

\item S. W. Hawking, {\elevenit Commun. Math. Phys.} {\elevenbf 43} (1975) 199.

\item J. D. Brown and J. W. York, {\elevenit Phys. Rev.} {\elevenbf D47} (1993)
1407.

\item J. D. Brown and J. W. York, {\elevenit Phys. Rev.} {\elevenbf D47} (1993)
1420.

\item J. W. York, {\elevenit Phys. Rev.} {\elevenbf D33} (1986) 2092.

\item J. D. Brown, E. A. Martinez, and J. W. York,
{\elevenit Ann. N. Y. Acad. Sci.} {\elevenbf 631} (1991) 225.

\item  J. D. Brown, E. A. Martinez, and J. W. York,
{\elevenit Phys. Rev. Lett.} {\elevenbf 66} (1991) 2281.

\item R. C. Tolman, {\elevenit Phys. Rev.} {\elevenbf 35} (1930) 904.

\item  H. W. Braden, J. D. Brown, B. F. Whiting, and J. W. York,
{\elevenit Phys. Rev.} {\elevenbf D42} (1990) 3376.

\item K. W. Howard, {\elevenit Phys. Rev.} {\elevenbf D30} (1984) 2532.

\item B. P. Jensen and A. C. Ottewill, {\elevenit Phys. Rev.} {\elevenbf D39}
(1989) 1130.

\item D. N. Page, {\elevenit Phys. Rev.} {\elevenbf D25} (1982) 1499.

\item D. Hochberg, T. W. Kephart, and J. W. York, {\elevenit Phys. Rev.}
{\elevenbf D48} (1993) 479.

\item M. R. Brown, A. C. Ottewill, and D. N. Page, {\elevenit Phys. Rev.}
{\elevenbf D33} (1986) 2840.

\item J. W. York, {\elevenit Phys. Rev.} {\elevenbf D31} (1985) 775.

\item D. Hochberg and T. W. Kephart, {\elevenit Phys Rev.}
{\elevenbf D47} (1993) 1465.

\item J. W. York, ``Entropy of a Conformal Scalar Field and a Black Hole"
(1985) unpublished.

\item G. L. Comer, ``The Thermodynamic Stability of Systems Containing Black
Holes",
University of North Carolina
doctoral thesis (1990) unpublished.
\end{list}
\end{document}